# Reversible nanopore sealing and *in situ* iron oxide nanoparticle synthesis on thin silicon nitride membranes

Zehui Xia[1*], Pia Bhatia[2*], Celia Morral[2*], Brian DiPaolo[1], Iryna Golovina[3], Adriana Buvač-Drndić[1], Chih-Yuan Lin[2], David Niedzwiecki[1], Marija Drndić[2]

[1.] *Goeppert LLC, 3401 Grays Ferry Avenue, Philadelphia, Pennsylvania 19146, United States*

[2.] *Department of Physics and Astronomy, University of Pennsylvania, Philadelphia, Pennsylvania 19104, United States*

[3.] *The Laboratory for Research on the Structure of Matter, University of Pennsylvania, Philadelphia, Pennsylvania 19104, United States*

\* *These authors contributed equally to the work.*

## Abstract

We report *in-situ* synthesis of iron oxide particles inside silicon nitride nanopores via a chemical reaction, monitored by current readout. Nanopores were formed by electroporation on glass chips (diameters from 1.7 to 11.3 nm), transmission electron microscopy (TEM) drilling (diameters from 6.5 to 64.6 nm), or hydrofluoric acid (HF) etching (diameters from 12.6 to 36.2 nm) in 5 to 20 nm thick membranes. Nanopores seal on timescales from ~1 ms to ~3.6 s, across a range of sizes and concentrations. We show single and ~5-pore arrays, as fabricated, after sealing, and after cleaning and pore recovery. These results are independent of fabrication method. Energy dispersive X-ray spectroscopy (EDS), aberration-corrected scanning TEM (AC-STEM), and powder X-ray diffraction (XRD) verify the synthesis of mixed magnetite and maghemite iron oxide. This work advances nanoparticle-nanopore chips for applications in biosensing, plasmonics and photonics when position and size control is required.

Solid-state nanopores are increasingly explored for biomolecule detection[1], water desalination[2,3], energy harvesting, and matter separation, among other uses[4]. An interesting use of solid-state pores is as nanoscale reactors for particle synthesis. The so-called "template synthesis" of nanostructures involves using nanoporous membranes to control the synthesis products[5] and has been demonstrated in track-etched polymer[6–9] and anodized aluminum oxide[6,7,10–15] membranes. *Venta et. al*[16] demonstrated electrically driven gold synthesis inside of single transmission electron microscopy (TEM) drilled silicon nitride ($SiN_x$) pores. In contrast to nanoporous membranes typically used for template synthesis, single nanopores in $SiN_x$ and 2D materials are fabricated in a deterministic fashion[17]. Such devices and their few-pore counterparts are ideal for biosensing, single-molecule detection[18], as well as in plasmonics[19–21] and photonics[22,23], where pore number, size, and location must be controlled. For example, a single nanoparticle next to a nanopore can serve as an integrated optical antenna[23] for plasmonics, and a single nanopore can be integrated within a field-effect transistor for high-bandwidth sensing[19,24,25].

Here, we demonstrate positional control of particle synthesis as well as reversible sealing of single $SiN_x$ pores and small arrays (4 to 5 pores) in 5 by 5 mm chips. Iron oxide particles inside nanopores are synthesized via a chemical reaction between ferric, ferrous and hydroxide ions in electrolyte solution, monitored by current readout. We reproduce the main features of the nanopore "sealing" and pore "recovery" in 18 $SiN_x$ membranes on glass and Si chips via *in situ* iron oxide synthesis **(Figure 1)**. From ionic current vs. time traces we characterize pore sealing time and dynamics. We also measure the resulting particle diameter for a range of precursor ($Fe^{2+}$, $Fe^{3+}$ and $OH^-$) concentrations, in low (50 mM KCl) and high (1M KCl) electrolyte concentration. Pore diameters were measured via TEM and/or estimated from the correlated ionic conductance. First, we discuss the $SiN_x$ pore fabrication using electroporation, TEM drilling, and hydrofluoric (HF) acid etching. Then we show how iron oxide synthesis was monitored in real time via ionic current suppression across the pore. We use TEM imaging, dispersive X-ray spectroscopy (EDS), aberration-corrected scanning TEM (AC-STEM) imaging, and powder X-ray diffraction (XRD) to verify the synthesis of iron oxide particles within pores and characterize their structure. Finally, we recover open pores by rinsing with diluted sulfuric acid ($H_2SO_4$) or piranha cleaning to dissolve the iron oxide.

The impact of this work spans several applications. First, it provides a means for reversible nanopore sealing. Nanopores can be sealed for storage and shipment and subsequently opened and cleaned. This is important for single nanopore sensors wherein the particles serve as "caps" that can be removed prior to measurement. A second benefit is in applications where the *precise placement* of few to single nanoparticles on a chip is beneficial – in contrast to nanoporous membranes with high densities of pores[5,6,26]. Iron oxide nanoparticles are candidate materials to enhance electronic transport in 2D materials and to fabricate magnetic superlattices[27]. Finally, electrical monitoring of nanoparticle growth can be exploited for the partial and sequential growth of composite particles beyond iron oxide[28] and functionalization of nanopores with specific materials at their walls[29].

To demonstrate the versatility of this approach, we employed TEM drilling, HF acid etching[30], and electroporation[18] to fabricate $SiN_x$ pores (**Figure 1a**). TEM drilling is generally considered the gold standard for nanopore fabrication[31,32]. Electroporation is rapid and inexpensive. Here we developed a simple passive-shunt resistor electroporation for low noise glass chips[33] featuring significantly smaller resistance-capacitance (RC) constants (**Figures S1-S2**). Single pores were made in membranes ~ 20 nm thick. Small nanopore arrays were fabricated in ultra-thin (~ 5 nm thick) localized circular regions patterned onto $SiN_x$ membranes. In this case,

SiN$_x$ membranes were pre-patterned by electron beam lithography and thinned by reactive ion etching (Oxford Instruments Plasma Lab 80+ RIE) to form locally thinned trenches (diameter ~ 100 nm)[34]. Nanopores were opened in these trenches by HF etching[30]. In all cases, pore formation was confirmed electrically (linear current-voltage (*I-V*), conductance-derived diameter estimate, stable open pore current) and iron oxide particle formation is *not* contingent upon the fabrication protocol.

**Figure 1b** shows SiN$_x$ nanopore sealing by iron oxide. Sealing was verified by *I-V* measurements, TEM, or both. We demonstrate iron oxide synthesis inside twelve TEM-drilled nanopores (**T1-T3, P2, P4, P6, P8, P10-13,** and **P15**), four electroporated nanopores (**E1-E4,**) and two nanopore arrays (arrays **A1** and **A2**).[*] Prior to particle formation, we measured the *I-V* curve of each device using 50 mM or 1 M KCl solution and estimated the effective pore diameter from ionic conductance ($d_{pore,\ before\ sealing}$) using **Eq. (S1)** (**Table S1**). We show *I-V* curves of two representative devices (pore **E2** and array **A1**) in **Figures 2a** and **2c** and their corresponding real-time current traces during pore sealing in **Figures 2b** and **2d**, respectively.

To initiate the synthesis reaction, NaOH was added to the *trans* side the nanopore; FeCl$_3$ and FeCl$_2$ (molar ratio of 2:1) were added to the *cis* side. All reagents were dissolved in unbuffered 50 mM or 1 M KCl. Iron oxide (Fe$_3$O$_4$) particles likely formed via the reaction[38-41]:

$$Fe^{2+} + 2\,Fe^{3+} + 8\,OH^- \rightarrow Fe_3O_4(s) + 4\,H_2O \qquad (1)$$

When the pore was sealed, the reagents could no longer mix, and ionic current fell to almost zero. We subsequently replaced the reagent solutions with KCl to confirm that there was little to no ion flow across the sealed pore (**Figures 2a** and **2c, green circles**). Using this current trace, we also estimated $d_{pore,\ after\ sealing}$ after pore sealing (**Table S1**) using **Eq. (S1)** [†]. In a few devices, ionic current fell substantially but did not reach 0. Any residual current was used to estimate an effective diameter $d_{pore,\ after\ sealing}$, corresponding to the pore area still accessible for ion transport. In all cases, $d_{pore,\ after\ sealing} \ll d_{pore,\ before\ sealing}$.

Pore **E2** in **Figure 2b** was fabricated by electroporation with a $d_{pore,\ before\ sealing}$ = 5.6 nm. In the first ~ 15 s, both the *trans* and *cis* sides of a flow cell were filled with KCl solutions, and $I \approx$ - 7 nA at V = - 300 mV (shown as **1. KCl** in **Figure 2b**). At t = 40 s, while keeping V = - 300 mV, KCl in the *trans* side was replaced with NaOH (in KCl), and $I \approx$ - 5.5 nA (**2. OH⁻ added** in **Figure 2b**). There was a non-zero current shift due to the chemical gradient created by the reagents[16]. At t = 60 s, KCl in the *cis* side was replaced with $FeCl_2$ and $FeCl_3$ (in KCl) and the current dropped to - 15 nA (**3. Fe²⁺, Fe³⁺ added** in **Figure 2b**). The voltage was reversed to V = + 500 mV at t = 70 s. This was accompanied by a sharp upward current spike, due to the capacitive response of the system[16]. After that, ionic current continuously fell until reaching a plateau (**4. Particle formed and pore sealed** in **Figure 2b**).

We also observed that iron oxide synthesis could proceed via reactant diffusional mixing *without* applied voltages. This is demonstrated in nanopore array **A1** (**Figure 2d**). NaOH and $FeCl_2$, $FeCl_3$ were added to the flow cell. Current gradually approached zero, even with an applied bias (+10 mV). This suggested that reagents approached the pore primarily by diffusional mixing, rather than by electroosmotic and/or electrophoretic forces[39]. Therefore, we observe that iron oxide synthesis inside $SiN_x$ pores does not necessarily require applied voltages.

---

[†] For nanopore arrays, this estimated diameter represents the effective pore diameter if the array was replaced with a single pore.

To directly confirm particle formation, we include representative TEM images and EDS mapping[‡] of pore **P10**. **Figures 2e-2g** show bright-field TEM images of pore **P10** before sealing, after sealing, and after recovery. **Figure 2h** is an annular dark-field (ADF) TEM image of the sealed pore corresponding to the Fe and O elemental maps in **Figures 2i** and **2j,** respectively.

**Figure 2k** displays the TEM-measured particle diameter, $d_{particle}$, *vs*. the pore diameter, $d_{pore}$, for single pores and for individual pores within two 3 by 3 arrays. **Figure 2k** also includes datapoints using single nanopore diameters estimated from the ionic conductance (open symbols in **Figure 2k**) and they lie close to the points using TEM-measured diameters. This good agreement from two correlated measurements validates the diameter values.

For single pore chips, most datapoints lie on or slightly above the line $d_{particle} = d_{pore}$, indicating that $d_{pore} \leq d_{particle}$. For the pores within the arrays (array **A1** and **A2**), the particle diameters are typically ~ 10 times larger than pore diameters. Data suggest that 50 mM KCl (used in array experiments) yields larger particles than 1 M KCl (used for single-pore chips), and this warrants a systematic ionic-strength investigation in the future.

We recovered the pores by dissolving the particles in 0.38 M $H_2SO_4$ for 10 minutes (pore **E2**) or piranha cleaning for 10 minutes (array **A1**) and rinsed the chip with water three times (**Figure 1c**).

The iron oxide nanoparticles can be removed via the reaction:

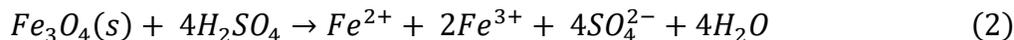

$$Fe_3O_4(s) + 4H_2SO_4 \rightarrow Fe^{2+} + 2Fe^{3+} + 4SO_4^{2-} + 4H_2O \qquad (2)$$

We re-measured the *I-V* curve (**Figure 2a, blue triangles**), to estimate the $d_{pore, recovered}$ after particle dissolution (**Table S1**). When comparing $d_{pore, recovered}$ with diameters before ($d_{pore, before sealing}$) and after ($d_{pore, after sealing}$) particle formation (**Table S1**), we found that $d_{pore, recovered}$ was slightly larger than $d_{pore, before sealing}$. This is probably due to the unintentional etching of nanopores after dissolving the iron oxide and measuring the $I$-$V$ in KCl solution[40]. To eliminate pore expansion, future experiments can include diluting reagents to remove the particles without etching the membrane, and/or coating the pores[40]. Furthermore, we observed the rectification in the recovered **E2** (**Figure 2a, blue triangles**) evidenced by the deviation from a linear, symmetric $I$-$V$ curve expected for an Ohmic pore. This is consistent with the residual iron oxide inside the pore after cleaning (**Figure 2g**), which modifies the pore surface charge and introduces asymmetry in the pore structure, thereby causing current rectification at a medium KCl concentration (50 mM KCl for blue triangles)[26].

---

[‡] A summary of EDS acquisition parameters is included in Table S2 in the *Supporting Information*.

We analyzed the sealed nanopores with TEM and EDS to verify the formation of particles and their composition. This was best demonstrated in array **A1** with locally thinned regions for easier pore identification. The array consists of a 3 by 3 array of nine ~ 100-nm-large, thinned (recessed) regions (~ 5 nm thickness) in the $SiN_x$ membrane, labeled from 1 to 9 (**Figure 3 (1a)**). Nanopores were fabricated by HF etching in thinned regions labeled 2, 4, 6, 8, and 9 (**Figures 3 (1c) to (1g)**). Regions labeled 1, 3, 5, and 7 remained intact as controls (**Figure 3 (1b)**). Due to the irregular pore shapes, pore areas were obtained using ImageJ: ~ 250, 236, 1030, 690, and 629 $nm^2$ for regions 2, 4, 6, 8, and 9, respectively (**Figure 3 (1c) to (1g)** and **Table S1**). We also include a 3D numerical simulation[41] demonstrating that a large separation ensures individual pores behaves independently with negligible interaction[42,43] (**Figure S11**).

TEM images of the sealed arrays and thinned regions *after* pore sealing are shown in **Figures 3 (2a)**. TEM images confirmed iron oxide formation as the thinned regions were sealed by particles and particle aggregates, corroborating $I$-$V$ measurements. Particles form *only* in the thinned regions with nanopores (regions 2, 4, 6, 8, 9, **Figures 3 (2c) to (2g)**), while the others (1, 3, 5, and 7) were *unchanged* from prior TEM imaging since they did not contain nanopores (**Figure 3 (2b)**).

The growth of iron oxide can also extend past the locally thinned region (**Figure 3 (2c) to (2g)**) if the reaction is long enough. We compared the pore area with the corresponding iron oxide particle area in **Figure 3 (5)**. The larger the nanopore is, the larger the particle that occupies it and its surrounding membrane. We repeated the experiment with 100 times diluted precursors than in **Figure 2** (3.2 mM *vs.* 0.32 M for NaOH, 0.8 mM *vs.* 0.08 M FeCl₃ and 0.4 mM *vs.* 0.04 M for FeCl₂) using nanopore array **A2 (Figures S9-S10)** and found that the area of the iron oxide was smaller even for pores of greater sizes than in array **A1**. For example, in array **A1**, the pore area in region 2 was ~ 250 nm$^2$ and the corresponding iron oxide particle area was ~ 0.08 μm$^2$. In array **A2**, a larger pore (~ 378 nm$^2$) in region 1 resulted in a smaller particle of ~ 0.016 μm$^2$ (**Figure S9** and **Table S1**). Further experiments can include tuning the precursors, varying pore sizes and spacings, and controlling the current to restrict particle formation. Partial nanopore filling can be of interest to create pores with desired properties, as well as modulate membrane porosity. *In situ* liquid cell TEM may be employed to observe reaction dynamics within the pore[44–47].

We used EDS to confirm the elemental composition of the synthesized particles. In **Figure 3 (4)**, we show ADF images of five nanopores (numbered 2, 4, 6, 8 and 9 in **Figure 3 (1a)**) in **A1** that were covered by particles (**Figure 3 (2a)**). The corresponding EDS maps of iron (**Figure 3 (4b)**) and oxygen (**Figure 3 (4c)**) confirm the synthesis of iron oxide. Strong Fe and O signals *exclusively* occur across the pore and thinned regions, confirming the localized particle formation.

We further imaged array **A1** to obtain EDS maps *after* the removal of the particles by piranha solution and the recovery of nanopores (**Figure 3 (3a)**). Little to no particle residue in **Figure 3 (2)** was observed in the recovered array and corresponding single nanopores in **Figure 3**

**(3)**. The ADF image of the same array in **Figure 3 (4d)** and the weak EDS signals from Fe in **Figure 3 (4e)** confirmed the removal of iron oxide. The widespread weak signal from oxygen across the array and the membrane (**Figure 3 (4f)**) can be from a native silicon oxide[40], also observed in recovered nanopores in **Figures 3 (3c)** to **(3g)**.

To examine the crystal structure of the iron oxide, AC-STEM imaging was performed on a ~ $20 \times 20$ nm$^2$ area in the center of particle 4 in array **A1 (Figure 4a** and **4b)**. Possible phases include crystalline magnetite ($Fe_3O_4$), maghemite ($\gamma$-$Fe_2O_3$) or a combination of both phases based on our synthesis involving the coprecipitation of $Fe^{2+}$ and $Fe^{3+}$ in the presence of $OH^-$ [37,48,49]. AC-STEM images revealed the polycrystalline nature of the particles **(Figure 4c** to **4e)**. The inset Fast Fourier transforms emphasize the non-uniform crystal orientations. A characteristic d-spacing was obtained by plotting an intensity profile from one of the atomic resolution images (**Figure 4e** and **4f)** to determine the iron oxide phase. This intensity profile was Fourier transformed to produce the plot in **Figure 4g.** The first peak at ~ 3.4 nm$^{-1}$ corresponds to a spacing of ~ 2.9 Å between atomic planes. Magnetite and maghemite both exhibit d-spacing near ~ 2.9 Å located at 2.967 Å and 2.95 Å, as determined by powder XRD, respectively[37,50]. Because magnetite can be oxidized to maghemite, a mixture of the two is likely present[36,49,51,52]. Another means of differentiating magnetite from maghemite is by visual inspection. Magnetite appears black to the eye whereas maghemite is a reddish-brown[53]. In our case, the precipitates appear blackish brown, suggesting that both phases are present (**Figure S12).** Powder XRD data corroborated this hypothesis, confirming the presence of magnetite and maghemite (**Figure 4h**). Additional investigation into the phases of iron oxide formed could utilize Mössbauer spectroscopy[37], Raman spectroscopy[48,49,54], and/or electron energy loss spectroscopy[55] to characterize these nanoparticles.

To illustrate the effect of precursor concentration, **Figure 5** compares results from two 3 by 3 arrays (**A1** and **A2**) of thinned $SiN_x$ regions (light grey areas in **Figure 5A**) containing 5 and 4 pores, respectively. Nanopore regions are encircled. The precursor concentration for array **A2** was 100 times smaller than for **A1**, resulting in $\sim 2.5$ times smaller particles on average, for similar-size pores. Estimating from TEM, the average particle size in **A1** was $\sim 370$ nm *vs.* $\sim 140$ nm for **A2 (Table S1)**.

We also investigated particle size and pore sealing time for single-pore and array chips. The growth is monitored as the current decays gradually to almost zero and is well fitted by a single exponential $I(t) \sim \exp\left(-t/\tau\right)$ (**Figure 5B**), where $\tau$ is the characteristic pore sealing time. Larger $\tau$ means slower sealing (more gradual current decay) while smaller $\tau$ suggests faster sealing. The $\tau$ values for single nanopores and arrays are summarized in **Figures 5C** and **5**D as a function of the initial pore diameter and the final particle diameter; $\tau$ ranges from $\sim 0.001$ s to $\sim 3.6$ s (**Table S1**). Overall, $\tau$ increases with the TEM-measured $d_{\text{particle}}$ for both single-pore and pore-array-chips, meaning that larger particles grow over a longer time. Together with **Figure 2k**, these data provide insights into characteristic sealing times as a function of nanopore and particle diameters from 8.5 nm to 64.5 nm, and 7.1 nm to 441.1 nm, respectively. Collectively, these data reveal sample-to-sample differences as we observe aggregation/clustering, variable particle uniformity, and cases where the particle diameter exceeds the pore diameter. A range of parameters can be adjusted in the pore sealing process to further control size and properties of particles. Future work can attempt to systematically vary the pore aspect ratios and electrolyte concentration. We presented a rapid current-monitored method for silicon nitride nanopore sealing allowing for easy pore recovery for biosensing experiments.

In conclusion, we demonstrated reversible sealing of single nanopores and few-nanopore arrays in $SiN_x$ using iron oxide particles. This work has relevance for both nanopore and nanoparticle-based technologies. Nanoparticle formation and pore sealing was monitored by ion current suppression, from the characteristic open pore current to almost zero. Synthesis was independent of the nanopore fabrication methods used and particles could be dissolved to recover the original pores. We characterized iron oxide particles revealing the presence of magnetite and maghemite. Future work could further probe particle formation and properties as a function of pore diameter, thickness and shape by sequential TEM imaging at different stages of pore sealing and by *in situ* liquid TEM imaging[56]. Given our observations of gradual pore sealing, $H_2SO_4$/piranha treatment reversibility and asymmetrical *I-V* curve rectification, the particle formation mechanism could be compatible with multiple mechanistic pathways. Iron oxide can nucleate in the pore interior, adsorb onto pore walls and continue growing *or* nucleate in the pore interior and grow freely until the pore wall begins to constrict growth[57,58]. Future work can explore these details by using in situ imaging/EDS and surface coatings to localize precursors, and develop models of particle growth across various pH, ionic strength, and concentrations. Our work advances particle synthesis to the single-pore regime and in pore arrays fabricated by top-down approaches, thereby opening the door to a range of advanced on-chip applications in biosensing, plasmonics and photonics.

## Supporting Information

The Supporting Information is available free of charge on the ACS Publication website.

## Author Information

*Corresponding Author(s)*

*Email: drndic@physics.upenn.edu, zx@gppert.com

*Author Contributions*

B.D., D.N. and Z.X. developed electroporation. Z.X. and B.D. performed electroporation, ionic testing, and DNA translocation measurements. Z.X. and C.M. performed nanopore sealing, and P.B. and C.M. performed TEM, AC-STEM and EDS. Z.X., P.B., C.M. A.B-D. and M.D. performed ionic and image analysis. D.N. and P.B. performed chip fabrication. C.Y.L. performed COMSOL simulations. I.G. performed powder XRD. All authors discussed the manuscript and contributed to the writing.

**Acknowledgements**

The work was supported by NASA STTR Phase I 80NSSC21C0368 and II 80NSSC23CA014 awarded to Goeppert LLC. P.B., C.M., C.Y. L. and M. D. were supported by NIH grants R21HG012395 and R01HG012413. CM acknowledges funding by NSF Graduate Research Fellowship Program under Grant No(s) (DGE-2236662). This work was carried out in part at the Singh Center for Nanotechnology, supported by the NSF grant NNCI-2025608. The authors acknowledge the use of facilities supported by NSF through the University of Pennsylvania MRSEC Center DMR-2309043. We acknowledge Mr. Andre Scott for nanopore chip fabrication and measurement, and Dr. Douglas Yates for assistance with TEM.

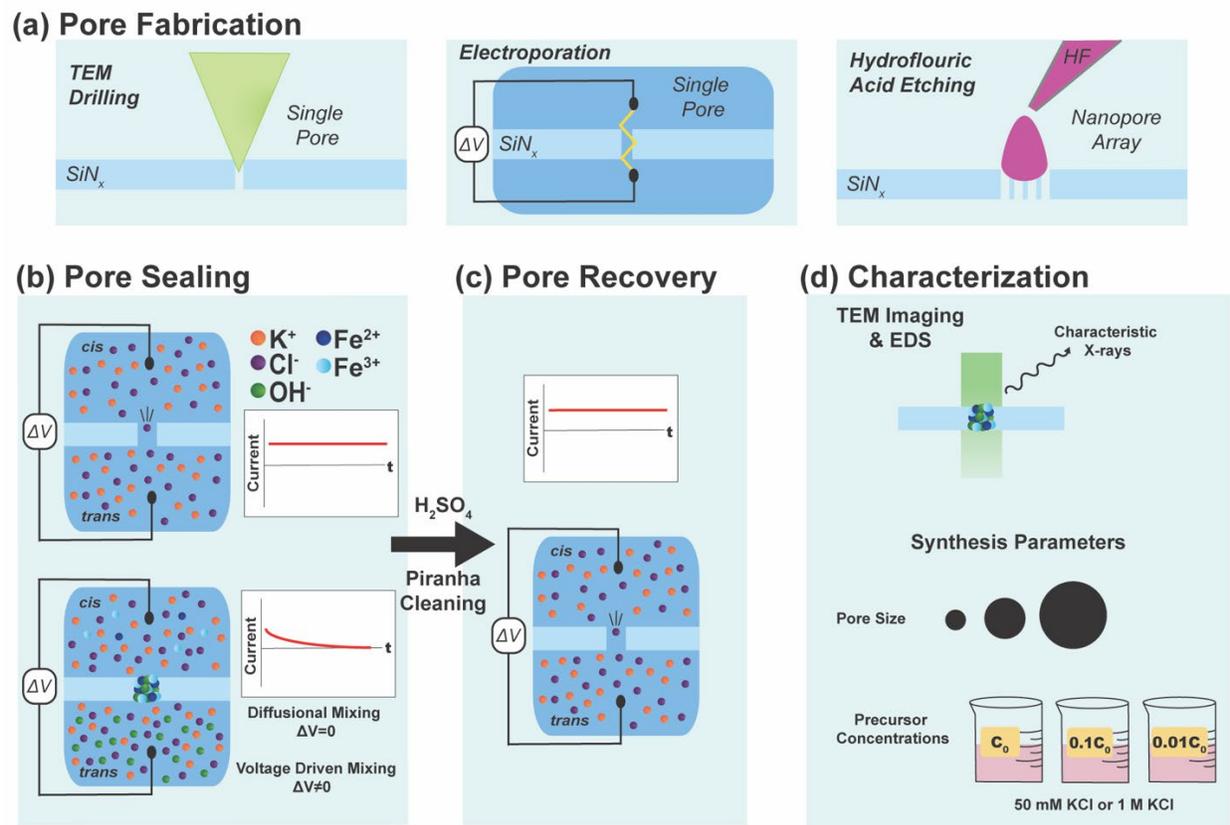

**Figure 1. Schematics of pore fabrication, pore sealing, pore recovery, and characterization methods**. **(a)** Pores and arrays were formed by either TEM drilling, electroporation, or hydrofluoric acid etching. **(b)** Iron oxide particle formation and pore sealing is initiated by adding NaOH to the *trans* side and $FeCl_3$ and $FeCl_2$ (molar ratio of 2:1) to the *cis* side. *Trans* and *cis* refer to the chambers in which the electrode is placed. This process is monitored in real time by measuring changes in ion current through the pore. The iron oxide particles grow within the nanopore until partially or fully sealing it. Sealing occurs with zero and non-zero applied voltages. **(c)** Particles can be dissolved with sulfuric acid or piranha, recovering the nanopore again for ionic measurements. **(d)** The pores and iron oxide particles were characterized via direct TEM imaging and energy dispersive x-ray spectroscopy (EDS). Parameters varied include nanopore size, electrolyte concentration and precursor concentration ($C_0$, *0.1C₀* and *0.01C₀*).

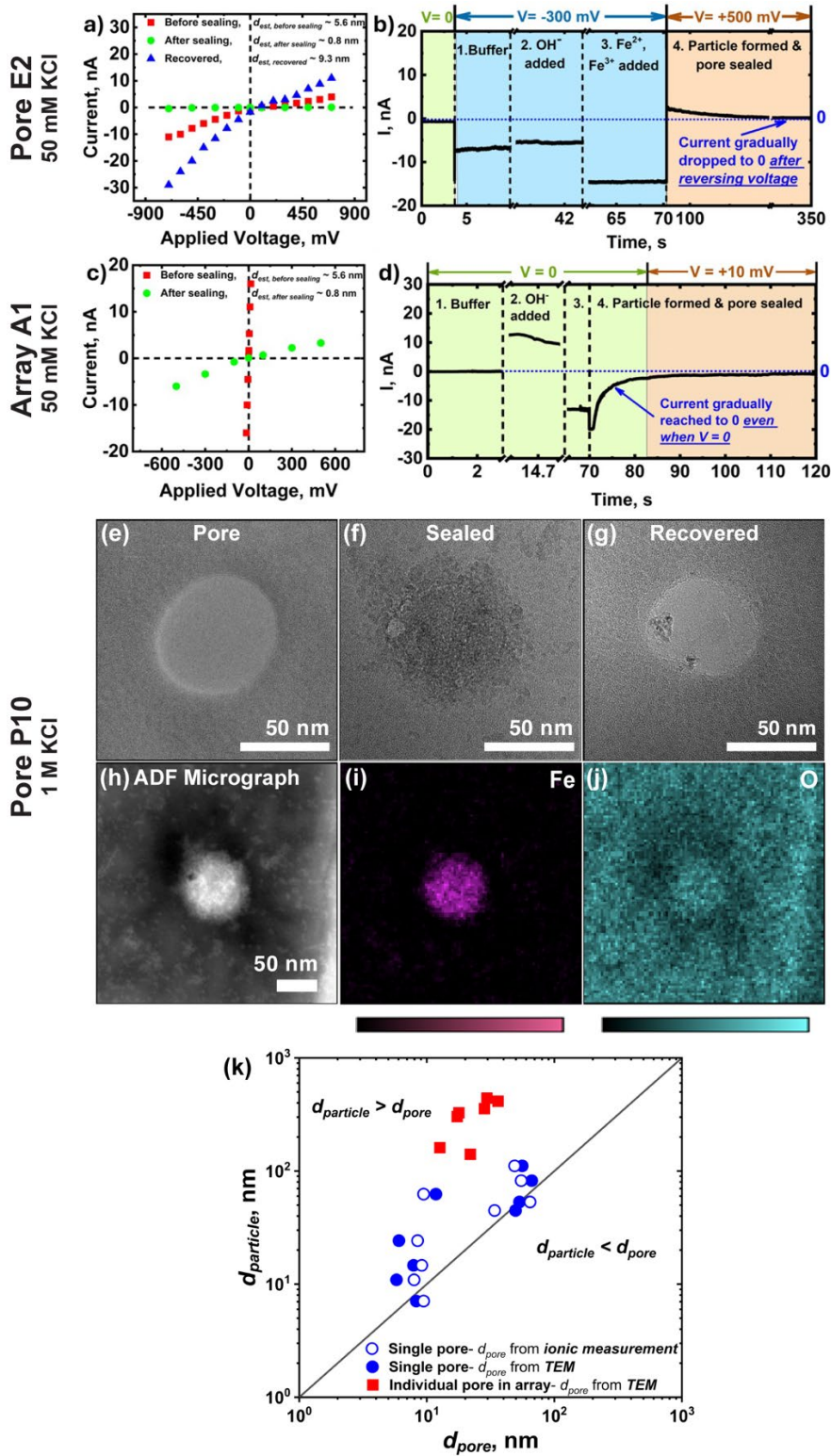

**Figure 2**. **Synthesis of iron oxide inside silicon nitride pores.** (**a-d**) Electrical monitoring of particle growth. *I-V* curves of two representative devices (Pore **E2** and array **A1**) and their corresponding real-time current traces during iron oxide nanoparticle formation and pore sealing. (**a**) Current-voltage (*I-V*) traces for Pore **E2** (fabricated by electroporation) before particle formation (red squares), after particle formation and pore sealing (green circles), and after recovery and particle removal in sulfuric acid (blue squares). (**b**) Real-time current *vs.* time trace and corresponding voltage during the iron oxide nanoparticle formation. Steps include, 1. 50 mM KCl was added to both sides, 2. 0.32 M NaOH was added to the *trans* side, 3. 0.08 M $FeCl_3$ and 0.04 M $FeCl_2$ (molar ratio of 2:1) was added to the *cis* side, and 4. voltage was then reversed to drive the reagents across the pore and iron oxide nanoparticle forms and seals the nanopore gradually. (**c**) Current-voltage (*I-V*) traces for array **A1** (nanopore array fabricated by HF etching) before (red squares) and after particle formation and pore sealing (green circles). (**d**) Real-time current *vs.* time trace and corresponding voltage during the iron oxide particle formation. Steps include, 1. 50 mM KCl was added to both sides, 2. 0.32 M NaOH was added to the *trans* side, 3. 0.08 M $FeCl_3$ and 0.04 M $FeCl_2$ (molar ratio of 2:1) was added to the *cis* side, and 4. voltage of +10 mV was then applied to check pore sealing. (**e-j**) TEM images of pore **P10**. Pore was sealed with no applied voltage using a precursor concentration of 0.8 mM $FeCl_3$, 0.4 mM $FeCl_2$ and 3.2 mM NaOH (*0.01$C_0$*) in 1 M KCl. (**e-g**) Bright field TEM images of Pore **P10** before sealing (**e**, $d_{pore,\ before\ sealing}$ ~55 nm), after sealing (**f**, $d_{pore,\ after\ sealing}$ ~1.9 nm) and after recovery (**g**). (**h**) Annular dark field (ADF) STEM images of the sealed nanopore. (**i-j**) Normalized EDS maps (Fe and O) after pore sealing. (**k**) Scatter plot of the TEM-measured particle diameter $d_{particle}$ *vs.* pore diameter $d_{pore}$ from both single nanopores and individual pores in array performed at zero applied voltage. Pore diameters were estimated from either ionic measurements or TEM images. The diagonal line $d_{particle} = d_{pore}$ is plotted and points above (below) the line indicate particles with diameters larger (smaller) than the pore.

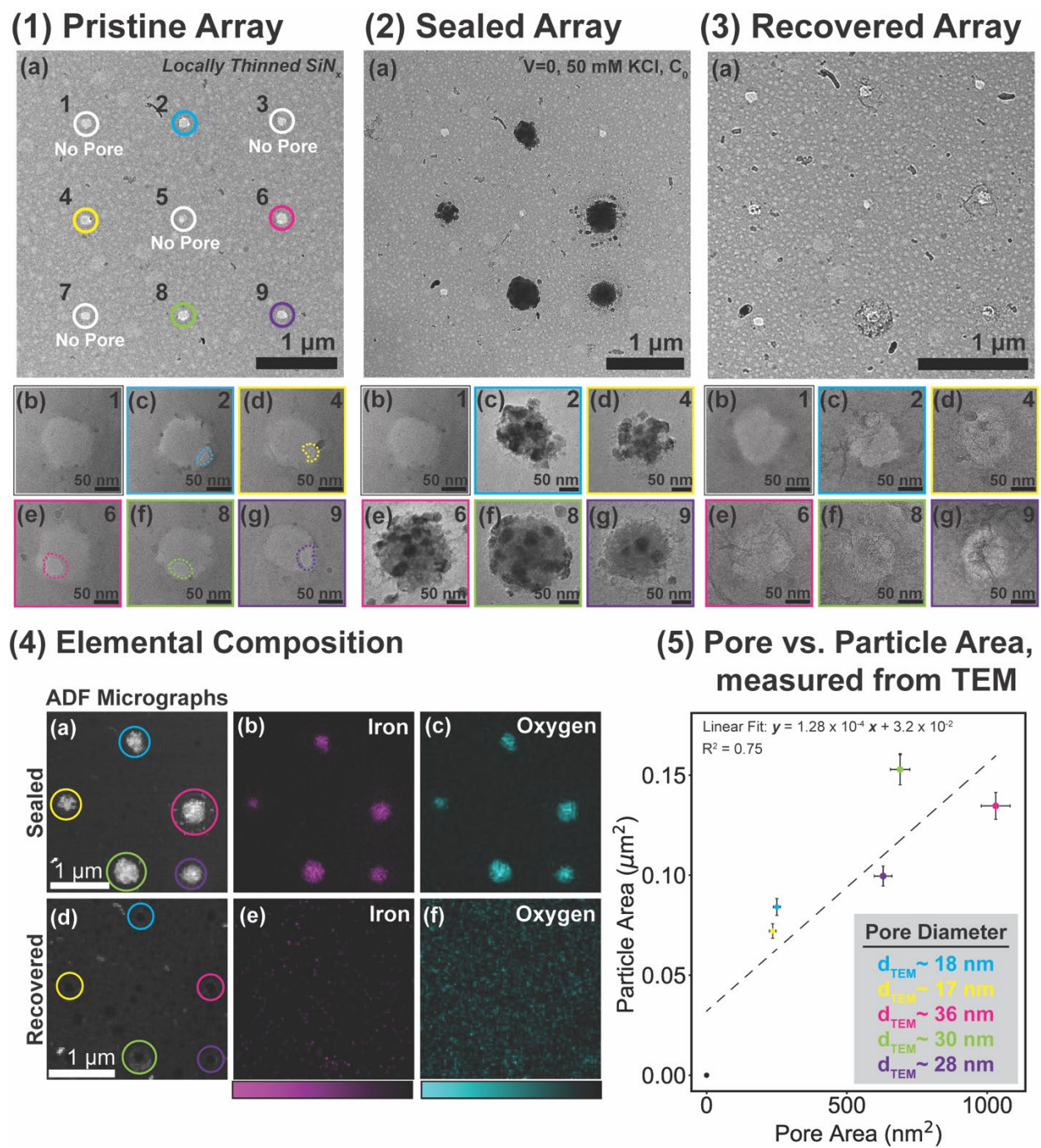

**Figure 3**. **Synthesis of iron oxide in nanopore array A1 and subsequent recovery. (1)** Pristine nanopore array *before* sealing. **(1a)** TEM image of a 3×3 array of ~ 100 nm large, locally thinned regions (~ 5 nm thick), patterned by electron beam lithography and thinned by RIE, numbered 1 - 9. Locally thinned regions are circled. Colored circles correspond to locally thinned regions which contain pores. White circles correspond to locally thinned regions without pores. Note that the same colors (blue, yellow, pink, green and purple) are used throughout the figure to identify the same locally thinned regions over time (2, 4, 6, 8, 9). (**1b-1g**) Higher magnification TEM images of the locally thinned regions 1, 2, 4, 6, 8, and 9, respectively. Nanopores inside of these thinned regions are outlined with colored dashed lines. (**2**) Nanopore array *after* sealing. (**2a**) TEM image of the 3 × 3 array after sealing with iron oxide particles. Iron oxide only fills the locally thinned regions with pores (regions 2, 4, 6, 8, and 9). (**2b-2g**) Higher magnification TEM images of locally thinned regions 1, 2, 4, 6, 8, and 9 after sealing with iron oxide. These images confirm complete occlusion of the pores shown in 1b-g. (**3**) Nanopore array *after* recovery. (**3a**) TEM image of the 3 × 3 array after piranha cleaning to dissolve the iron oxide. (**3b-c**) Higher magnification TEM images of locally thinned regions 1, 2, 4, 6, 8 and 9 after piranha cleaning. (**4**) Annular dark field (ADF) STEM images showing the sealed nanopore array and the elemental composition before and after recovery. (**4a**) ADF STEM image showing the sealed nanopore array. Regions 2, 4, 6, 8 and 9 are circled in different colors (blue, yellow, pink, green, and purple). (**4b-4c**) Normalized EDS maps corresponding to the ADF image shown in **4a**. These maps confirm the presence of iron and oxygen. (**4d**) ADF STEM image showing the recovered nanopore array after piranha cleaning. (**4e-4f**) EDS spectrum image maps corresponding to the ADF image shown in **4d**. These maps confirm the removal of iron oxide particles. (**5**) Plot and linear least squares fit of initial pore area versus resulting particle area. Pore areas were measured in ImageJ from images **1b-1g**. Particle areas were measured in ImageJ (**2b-2g**). Error associated with these measurements was approximated to ± 5% of the area. This plot shows that the particle size increases with pore size. The coefficient of determination ($R^2 = 0.75$) indicates a good fit between the experimental data and the linear regression model, $y = 1.28 \times 10^{-4} x + 3.2 \times 10^{-2}$.

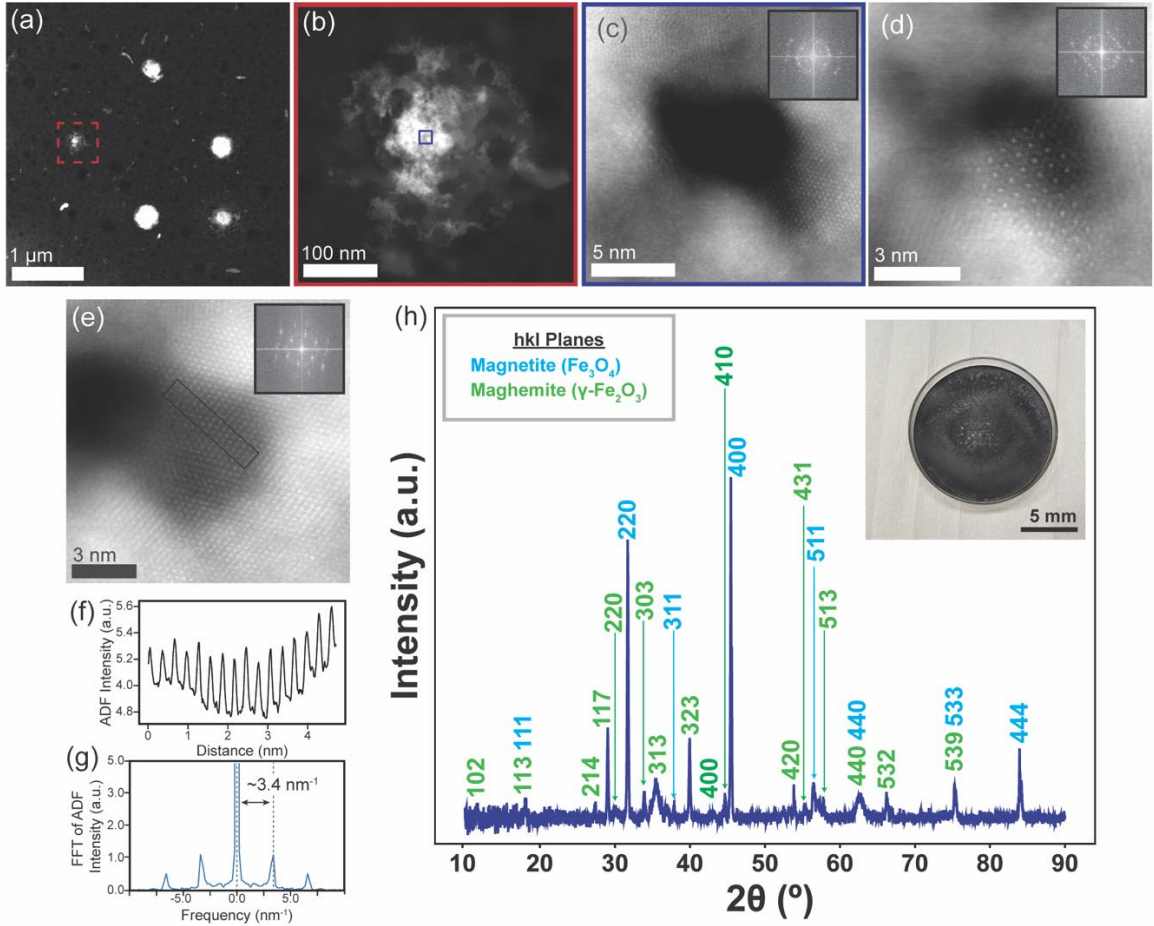

**Figure 4. AC-STEM and XRD characterization of iron oxide crystal structure in array A1.**
(**a**) ADF image of nanopore array **A1** after sealing the pores with iron oxide. The red dashed box corresponds to the zoomed-in image shown to the right. (**b**) Higher magnification ADF image of the iron oxide particle in region 4 of array **A1**. The blue box in the center of the image corresponds to the region shown in c. (**c**) Atomic resolution image of the center of the iron oxide formed in region 4, showing multiple crystal grain orientations around the black region in the center of the image with inset FFT. (Note that the black region in the center simply corresponds to a thinner region of the iron oxide particle.) (**d**) Another region of the iron oxide, nearby the region shown in (**c**), with inset FFT. This atomic resolution image emphasizes the differing crystal grain orientations present. (**f**) A third region of the iron oxide, nearby the regions shown in (**c**) and (**d**), with inset FFT. This region is primarily covered by a single crystal grain orientation. An intensity profile was taken from the region boxed in white to produce (**f**). The intensity plot was Fourier transformed to produce (**g**). The first peak occurs at ~ 3.4 nm$^{-1}$, corresponding to a d-spacing of ~ 2.9 nm. (**h**) Powder XRD spectrum of iron oxide precipitates. All diffraction peaks correspond to the phases of magnetite (hkl planes labelled in blue) and maghemite (hkl planes labelled in green). The peaks were identified using the International Centre for Diffraction Data (ICDD) PDF cards #01-082-3507 [59] and #01-089-5894 [60], respectively. The inset image shows the sample prior to measurement.

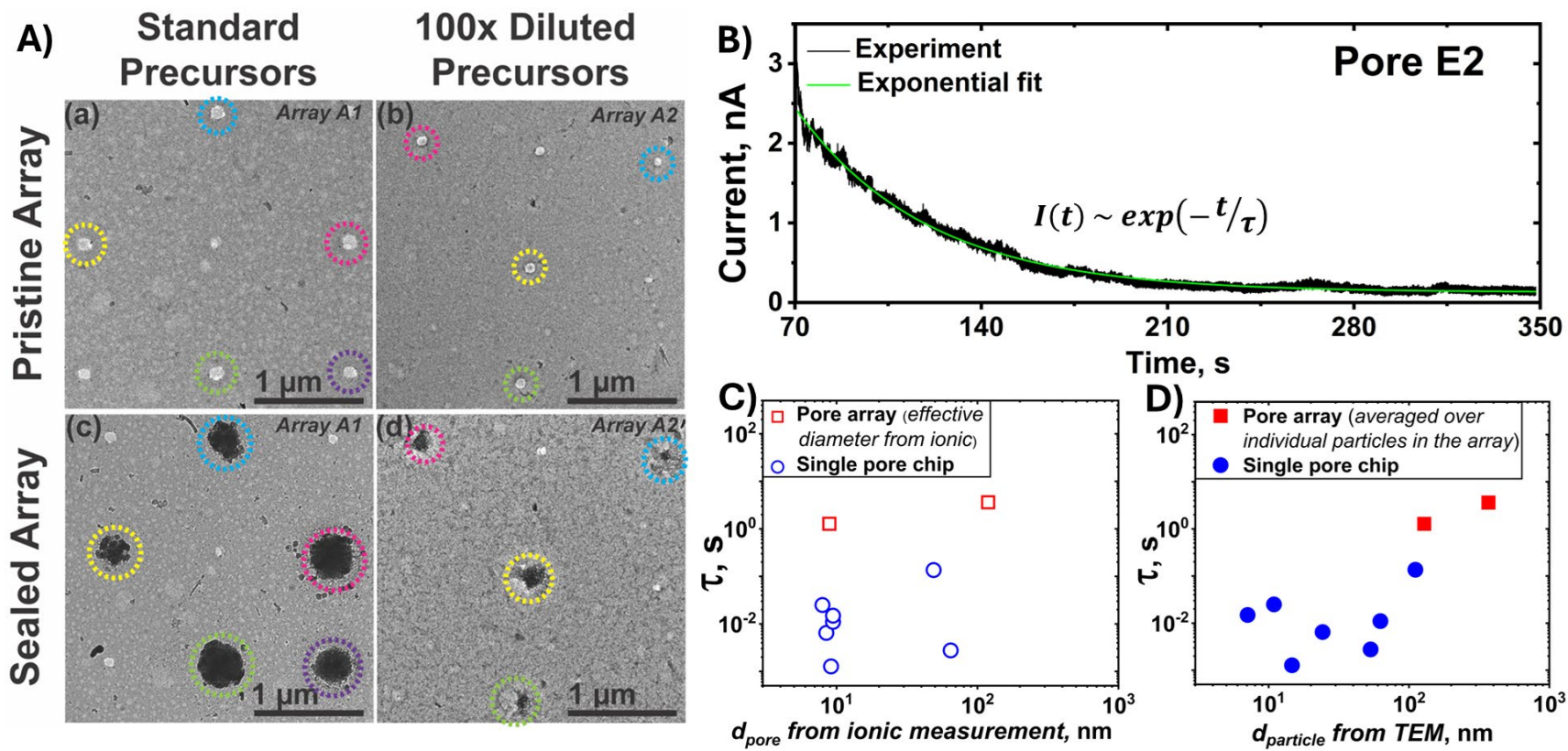

**Figure 5. Iron nanoparticle growth: kinetics, concentration and size dependence. A) (a-b) Qualitative comparison of the effect of precursor concentration on particle formation in nanopore arrays A1 and A2.** TEM micrographs of pristine, locally thinned arrays patterned into $SiN_x$ with electron beam lithography and RIE. The colored circles indicate locally thinned regions with pores. (**c-d**) TEM micrographs *after* sealing arrays **A1** and **A2** with iron oxide. To seal array **A1**, 0.04 M $FeCl_2$, 0.08 M $FeCl_3$, and 0.32 M NaOH ($C_0$) were used in 50 mM KCl. To seal array **A2**, 0.4 mM $FeCl_2$, 0.8 mM $FeCl_3$, and 3.2 mM NaOH (0.01 $C_0$) were used in 50 mM KCl.

The iron oxide particles formed in array **A1** extend beyond the nanopore and locally thinned regions onto the bulk membrane. The iron oxide particles formed in array **A2** are much smaller. **B)** Representative current *vs.* time trace from **Pore E2** shows that the current decays toward baseline and is well fit by a single exponential $I(t) \sim \exp\left(-t/\tau\right)$ over 70-350 s from **Figure 2b**. $\tau$ is the characteristic time scale of sealing the pore. Larger $\tau$ means slower sealing (more gradual decay) while smaller $\tau$ suggests faster sealing. **C)** Scatter plot of $\boldsymbol{\tau}$ *vs.* effective pore diameter estimated from ionic measurement $d_{pore}$ for nanopore array and single pore chips. Arrays exhibit longer sealing times than single pores at comparable "effective diameters" (a diameter of an equivalent single pore with the same conductance). **D)** Scatter plot of $\boldsymbol{\tau}$ *vs.* particle diameter estimated from TEM $d_{particle}$ for nanopore array (averaged over individual pores in the array) and single pore chips. Arrays exhibit sealing times ~1-2 orders of magnitude larger than single pores at comparable $d_{particle}$.

For Table of Contents Only

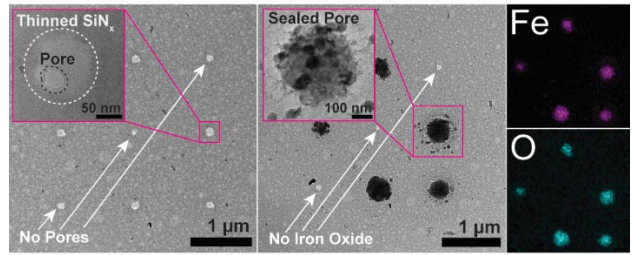